\documentclass[aps,pra,reprint,amsmath,amssymb,superscriptaddress]{revtex4-1}
\usepackage{graphicx}
\usepackage{mathrsfs}
\renewcommand{\phi}{\varphi}
\usepackage{xfrac}
\usepackage[colorlinks]{hyperref}
\usepackage{mathrsfs}
\usepackage{times}
\usepackage[utf8]{inputenc}
\mathchardef\Re="023C
\mathchardef\Im="023D
\usepackage{subcaption}
\usepackage{slashed}

\usepackage{color}

\makeindex

\begin{document}

\title{Rayleigh waves and cyclotron surface modes of gyroscopic metamaterials}

\author{Filip Marijanović}
\affiliation{Trinity College, University of Cambridge, Cambridge, United Kingdom}
\affiliation{Physik-Department, Technische Universität München, 85748 Garching, Germany}
\author{Sergej Moroz}
\affiliation{Physik-Department, Technische Universität München, 85748 Garching, Germany}
\affiliation{Munich Center for Quantum Science and Technology (MCQST), 80799 München, Germany}
\affiliation{Department of Engineering and Physics, Karlstad University, Karlstad, Sweden}
\author{Bhilahari Jeevanesan}
\affiliation{Physik-Department, Technische Universität München, 85748 Garching, Germany}
\affiliation{Munich Center for Quantum Science and Technology (MCQST), 80799 München, Germany}
\begin{abstract}

We investigate the elastic normal modes of two-dimensional media with broken time-reversal and parity symmetries due to a Lorentz term. Our starting point is an elasticity theory that captures the low-energy physics of a diverse range of systems such as gyroscopic metamaterials, skyrmion lattices in thin-film chiral magnets and certain Wigner crystals. By focusing on a circular disk geometry we analyze finite-size effects and study the low-frequency shape oscillations of the disk. We demonstrate the emergence of the Rayleigh surface waves from the bottom of the excitation spectrum and investigate how the curvature of the disk-boundary modifies their propagation at long wavelengths. Moreover, we discover a near-cyclotron-frequency wave that is almost completely localized at the boundary of the disk, but is distinct from the Rayleigh wave. It can be distinguished from the latter by a characteristic excitation pattern in a small region near the center of the disk. 
\end{abstract}

\maketitle

\section{Introduction}
The field of mechanical metamaterials has seen a lot of development lately, see reviews \cite{bertoldi2017flexible, huber2016topological}. Certainly part of the inspiration in this area stems from an interest in developing classical mechanics analogues of celebrated quantum mechanical systems \cite{ma2019topological}. A prominent example for this approach are gyroscopic metamaterials \cite{Nash14495,PhysRevB.97.100302,mitchell2021real,susstrunk2016classification,phononic_gyroscopes,garau2019transient,nieves2021directional}, which were developed as mechanical analogs of quantum Hall systems. The constant rotation rate of the gyroscopes around their axes implies an innately broken time-reversal-symmetry in these metamaterials. This provides the prerequisite for the existence of topologically protected edge modes. 
One way to view all of this activity is through the lens of effective field theory with the underlying symmetries serving as constraints. The very different settings, classical in one case and quantum in the other, give rise to similar physics, because the systems are governed by the same field theory at long wavelengths, with their forms severely constrained by symmetry.

In a previous work  \cite{2021rayleigh}, some of us explored a low-energy two-dimensional elasticity theory that is suitable for the description of systems with broken parity $P$ and time reversal $T$ symmetries, but intact combined $PT$ symmetry, originating from Lorentz forces. 
We have used this theory to explore the Rayleigh edge-wave phenomenology in elastic materials and found results that are applicable to platforms as diverse as gyroscopic metamaterials \cite{Nash14495, nieves2020rayleigh}, skyrmion lattices \cite{petrova2011spin,PhysRevLett.107.136804} and certain classes of Wigner crystals in magnetic fields \cite{DAmico1999a}. In that work our focus was on edge waves that propagate along a straight boundary of a semi-infinite plane. An immediate question, of practical importance in real-world applications, concerns the robustness of these predictions with regard to finite-size effects. What are the effects of the boundary curvature on propagating edge waves? 

After Rayleigh's seminal work \cite{Rayleigh1885} on surface modes localized near the boundary of a straight semi-infinite medium, the influence of a curved boundary was studied by Sezawa \cite{sezawa1927dispersion} and by Viktorov \cite{viktorov1967rayleigh}.  These authors considered propagation of Rayleigh-waves along cylinders and spheres. Given the shape of the Earth, the latter is a problem of practical interest in seismology and has therefore received thorough treatments in the literature, see e.g. \cite{ewing1969elastic} for a more detailed discussion of curvature effects. The principal finding in these works is that the curvature of the medium's boundary entails a modification of the Rayleigh-wave dispersion relation at low frequencies and small wavenumbers. 

Naturally, the results of these papers cannot be applied to the system of our interest since the Lorentz forces, which were not considered in these classic works, dominate the dynamics at low frequencies. One goal of the present paper is to address the consequences of finite system-size. 

We introduce the linearized elasticity theory in the presence of the Lorentz term in sec. \ref{sec:EFT} and find that the coupled partial differential equations for transverse and longitudinal displacements can be analytically solved in polar coordinates. We analyze the oscillation spectrum of the normal modes in the circular disk geometry in sec. \ref{sec:oscSpec}.  The section \ref{sec:NewSurfModes} contains several new results about surface modes.  Firstly, we explain how Rayleigh-waves fit into the oscillation spectrum. We find that the oscillation frequencies of Rayleigh waves at long wavelengths are exponentially diminished compared to the straight-boundary case \cite{2021rayleigh}, see Fig. \ref{fig:LogLogFitDisp} and eq. \eqref{eq:RayleighCrossover}. We also discuss the low-frequency shape oscillations of the disk and study its dependence on the Poisson ratio, see Fig. \ref{fig:pointosc}. Quite remarkably, we discover an additional branch of surface modes that emerge at frequencies close to the cyclotron frequency, see Fig. \ref{fig:surface}. This mode is almost fully localized at the boundary of the disk, except for a small region near the center.
The visual appearance of this mode is reminiscent of the famous Arago spot from wave optics, where diffraction generates a bright central spot in the shadow of a disk  \cite{sommerfeld1954lectures, hecht2017optics}. 
In the outlook \ref{sec:concl} we discuss experimental consequences for the observation of the cyclotron surface mode.

\section{Low-energy elasticity of media with Lorentz forces}\label{sec:EFT}
The starting point of our analysis is the elasticity theory of 
 two-dimensional media with broken $P$- and $T$- symmetries, but intact combined $PT$ symmetry. The degree of freedom of this model is a two-dimensional, coarse-grained field $\mathbf u(\mathbf x) = (u_x(\mathbf x),u_y(\mathbf x))$ describing small displacements of the medium from its equilibrium configuration. The theory is fully defined by the Lagrangian density 
\begin{equation}
\label{eq:Langrangian}
	\mathcal{L} = \frac{\rho }{2}[\partial_t{u(\mathbf x)}]^2 - \frac{\rho \Omega}{2} \epsilon_{ij} u^i(\mathbf x) \partial_t{u^j(\mathbf x)} - \mathcal{E}_{\text{el}}(u_{ij}(\mathbf x)),
\end{equation}	
where the first term is the kinetic energy of the displacement, with $\rho$ denoting the constant mass density of the medium. The second term is a Lorentz term that violates $P$- and $T$-symmetry, $\Omega$ is the cyclotron frequency and $\epsilon_{ij} $ the Levi-Civita tensor in two dimensions. The final term $\mathcal{E}_\text{el}$ is the elastic displacement energy density of the system, which is quadratic in the strain fields $u_{ij} = (\partial_i u _j + \partial_j u _i)/2$. The form of $\mathcal{E}_{\text{el}}$ is severely constrained by symmetry. In fact, imposing a six-fold rotation symmetry on a solid in two dimensions reduces its elastic energy density to the form \cite{landauElasticity}
\begin{eqnarray}
\mathcal{E}_\text{el} &\equiv& 2C_1 u^2_{kk} + 2 C_2 [u_{ij} - u_{kk} \delta_{ij}/2]^2 \\
&=& 2C_1 (\partial_k u^k)^2 + \frac{1}{2}C_2 [\partial_i u_j + \partial_j u_i - \partial_k u^k \delta_{ij}]^2,
\end{eqnarray}
with $C_1, C_2$, the compressional and shear moduli, being the only elastic parameters for this case.

The Lagrangian \eqref{eq:Langrangian} captures the low-energy physics of gyroscopic metamaterials \cite{Nash14495,PhysRevB.97.100302,mitchell2021real,susstrunk2016classification,phononic_gyroscopes, garau2019transient,nieves2021directional}. These are networks of elastically coupled gyroscopes that break time-reversal symmetry by spinning at a fixed rotation rate around their axis of symmetry. The Lorentz term dynamics is generated by the fact that a swiftly rotating gyroscope responds to an external force by moving transversely to it. In fact, our Lagrangian \eqref{eq:Langrangian} can be viewed as the continuum field theory governing the triangular-lattice models considered in  \cite{garau2019transient} at large length-scales and small nutation angles of the gyroscopes.

Another notable system which is described by Lagrangian \eqref{eq:Langrangian} is the skyrmion triangular crystal emerging in the thin-film chiral magnets of $\text{Fe}_{0.5}\text{Co}_{0.5}\text{Si}$ \cite{yu2010real} and FeGe \cite{yu2011near}.  A continuum field theory for these skyrmion lattices was worked out in the literature \cite{petrova2011spin,PhysRevLett.107.136804} and has precisely the form of eq. \eqref{eq:Langrangian}. The Lorentz term appears very naturally in this theory, since skyrmions in a ferromagnet experience an effective magnetic field $\mathcal{B}$, yielding a cyclotron frequency $\Omega = \mathcal{B}/m$, where $m$ is the mass of a skyrmion.

\subsection{Wave Motion in the Circular Disk Geometry}
Having introduced the elasticity theory that we wish to study, we will now analyze the oscillation modes of a circular disk of elastic material that is governed by the Lagrangian \eqref{eq:Langrangian}. We begin by working out the bulk eigenmodes in polar coordinates and then apply stress-free boundary conditions at the rim of the disk.

From the Lagrangian \eqref{eq:Langrangian} the equation of motion 
\begin{equation}
-\omega^2 u^i + i\omega\Omega \epsilon_{ij} u^j = 2v_1 \partial^i \partial_k u^k + v_2 \partial^j \partial_j u^i
\label{eq:eom}
\end{equation}
follows, where $v_i = C_i / \rho$ and we made the time-periodic ansatz $u^i(\mathbf r,t) = u^i(\mathbf r) e^{i\omega t}$. It is understood that the physical displacement field is obtained by taking the real part of $u^i(\mathbf r,t)$. 
As is customary \cite{thorne}, we solve this vector differential equation by decomposing the displacement field into a curl-free and a divergence-free part
\begin{equation}
\label{eq:decomposition}
u^i = \partial^i \psi + \epsilon^{ij} \partial_j \Lambda,
\end{equation}			
where $\psi$ and $\Lambda$ are scalar potentials. Inserting this decomposition into the equation of motion \eqref{eq:eom}, we can derive two independent equations by applying either $\partial_i$ or $\epsilon_{ij} \partial^j$ to it. 
Although at first sight these are fourth order equations, an overall trivial $\nabla^2$ operator can be removed from all the terms, yielding
\begin{eqnarray}
\label{eq:PDEpsiLambda1}
(2v_1 + v_2) \nabla^2 \psi + \omega^2 \psi + i \omega \Omega \Lambda &=& 0  \\
\label{eq:PDEpsiLambda2}
v_2 \nabla^2 \Lambda + \omega^2 \Lambda - i\omega\Omega \psi &=& 0.
\end{eqnarray}
In the absence of the Lorentz term, i.e. for $\Omega=0$, the two equations decouple into the usual longitudinal and transverse eigenmodes. However, the Lorentz term couples these modes such that the actual eigenmodes are superpositions of longitudinal and transverse waves. 

In order to solve the coupled system of Helmholtz equations \eqref{eq:PDEpsiLambda1} and \eqref{eq:PDEpsiLambda2} for the case of a circular disk, we switch to polar coordinates $(r,\theta)$ with the origin at the center of the disk. It is well-known that in polar coordinates the eigenmodes of the Laplacian operator are $J_n(q r) e^{i n\theta}$, where $J_n(x)$ denotes the $n$-th order Bessel function of the first kind and $q$ is the radial wavenumber, thus we have the identity
\begin{equation}
\nabla^2 [J_n(q r) e^{i n \theta}] = -q^2 J_n(q r) e^{i n \theta}.
\end{equation} 
Now any function $f(r,\theta)$ can be first expanded in a Fourier series $\sum_n f_n(r) \exp(i n \theta)$ and thereafter the functions $f_n(r)$ can be expressed as an integral over Bessel functions $\int_0^\infty dq q \hat f_n(q) J_n(q r)$, where $\hat f_n(q)$ is the Hankel transform of $f_n(r)$. Thus in the following it suffices to restrict our attention to $\psi$ and $\Lambda$ of the form
\begin{equation}
\label{eq:psiLambdaAnsatz}
\begin{split}
\psi &= A_n J_n(q r)e^{i n\theta}, \\
\Lambda &= i B_n J_n(q r) e^{in\theta}.
\end{split}
\end{equation} 
where the factor of $i$ in the second equation was introduced for a future convenience.
The other family of eigenmodes of the Laplacian are the Bessel functions of the second kind, which all have a singularity at $r=0$ and are therefore unphysical in this problem. We note that one can define the azimuthal wavenumber $k_\varphi \equiv n/R$ that is especially relevant for wave propagation along the disk boundary. We will use the term `wavenumber' both for $n/R$ and $n$, since $R$ will be a constant throughout this paper.

To simplify the notation, we now introduce the frequency scale
\begin{eqnarray}
\omega_0 \equiv \frac{\sqrt{2v_1 + v_2}}{R}
\end{eqnarray}
and measure the frequency $\omega$ and cyclotron frequency $\Omega$ in units of $\omega_0$
\begin{eqnarray}
f \equiv \frac{\omega}{\omega_0}, \hskip 5pt b \equiv \frac{\Omega}{\omega_0}.
\end{eqnarray}
According to eq. \eqref{eq:PDEpsiLambda1} the physical meaning of $\omega_0$ is that it sets the frequency scale for the longitudinal modes of the disk in the absence of the magnetic field ($\Omega = 0$). In the following, we will sometimes refer to $b$ as the (dimensionless) magnetic field.
We also measure the wavenumber $q$ in units of the inverse radius $R$ by introducing the dimensionless variable
\begin{eqnarray}
x \equiv q R.
\end{eqnarray}
With these simplifications, insertion of the ansatz \eqref{eq:psiLambdaAnsatz} into the wave equations \eqref{eq:PDEpsiLambda1}, \eqref{eq:PDEpsiLambda2} yields
\begin{equation}
\begin{pmatrix}
-x^2 +f^2 &  - f b\\
-\frac{2f b }{1-\sigma}& -x^2 + \frac{2f^2 }{1-\sigma}
\end{pmatrix} 
\begin{pmatrix}
A_n \\
B_n 
\end{pmatrix}						
 = 0,
\label{eq:matrix}
\end{equation}
where we expressed the ratio of the elastic constants $v_1, v_2$ through the Poisson ratio $\sigma \equiv {(2v_1 - v_2)}/{(2v_1 + v_2)}$. In order for this system of equations to have a solution, the coefficient matrix must have a vanishing determinant. This condition yields the dispersion relation in the form
\begin{equation}
 x^2_\pm = \frac{f^2}{2(1-\sigma)} \big[3-\sigma \pm (1+\sigma)\sqrt{1+8 \frac{b^2}{f^2} \cdot \frac{1-\sigma}{(1+\sigma)^2}}\big].
 \label{eq:disp}
\end{equation}
Thus the general solution for a given order $n$ and frequency $\omega$ is given by
\begin{eqnarray}
\psi &=& [A_n^+ J_n(q_+ r) + A_n^- J_n(q_- r) ] e^{i n\theta},
\label{eq:solnOrder1}\\
\Lambda &=& i [B_n^+ J_n(q_+ r) + B_n^- J_n(q_- r) ] e^{i n\theta}
\label{eq:solnOrder2}
\end{eqnarray}	
The ratio of amplitudes $A^{+}_n/B^{+}_n$ and $A^{-}_n/B^{-}_n$ is fixed by eq. \eqref{eq:matrix}, see eq. \eqref{eq:ratio} for the explicit formula.
Although the ratios $A^+/A^-$ and $B^+/B^-$ are arbitrary at this point, we will see below that boundary conditions fix them.

One of the immediate consequences of having a finite Lorentz term is that there exists a regime with $f<b$, where the frequency of oscillations is smaller than the cyclotron frequency. According to eq. \eqref{eq:disp}, $x_-$ is purely imaginary here and as a consequence the Bessel function carries an imaginary argument, implying a non-oscillatory radial decay of the $q_-$ solution. 
		
In order to fully define the wave propagation problem, we need to specify the boundary conditions at the edge of the disk $r=R$. These boundary conditions quantize the allowed values of $q_\pm$ and thus render the frequency spectrum discrete.   We make the assumption that the elastic medium is free at the boundary, in other words that no external forces act at the crystal on its outer surface. To implement this, we introduce the stress tensor $\mathbf T$ that is defined as a derivative of the elastic energy with respect to the strain field, $T_{ij} \equiv {\partial \mathcal E _\text{el}}/{\partial u^{ij}}$, thus
\begin{eqnarray}
T_{ij}/\rho = 2 v_1 u_{kk} \delta_{ij} + 2 v_2  [u_{ij} - u_{kk} \delta_{ij}/2].
\end{eqnarray}
The absence of external forces at the boundary implies $\mathbf T \cdot \mathbf n = 0$, where $\mathbf n$ is a normal-vector to the boundary of the disk. In polar coordinates this implies $T_{rr}=0$ and $T_{r\theta}=0$ and hence
\begin{eqnarray}
(2v_1+v_2)\partial_r u_r  + (2v_1 - v_2) \left[ \frac{u_r}{r} + \frac{1}{r}\partial_\theta u_\theta \right] &=&0, 
\label{eq:boundary1}\\
\partial_r u_\theta + \frac{1}{r} \partial_\theta u_r - \frac{u_\theta}{r}& =&0 
\label{eq:boundary2}
\end{eqnarray}	
at $r=R$.

These boundary conditions constrain the general solution \eqref{eq:solnOrder1}-\eqref{eq:solnOrder2} by allowing only a certain mixture of $+$ and $-$ modes, satisfying
		
\begin{equation}
\begin{pmatrix}
P_n(x_+,\alpha^+) &  P_n(x_-,\alpha^-)\\
Q_n(x_+,\alpha^+) &  Q_n(x_-,\alpha^-)
\end{pmatrix} 
\begin{pmatrix}
A^+ \\
A^-
\end{pmatrix}						
= 0 			 
\label{eq:det}
\end{equation}
where $P_n$ and $Q_n$ are functions involving Bessel functions and their derivatives, while $\alpha^\pm = B^\pm_n / A^\pm_n$ are the ratios of amplitudes of the $+$ and $-$ mode. The explicit forms of all these functions are lengthy and we therefore present them in the appendix \ref{app1}.
The system of equations \eqref{eq:det} can be satisfied if the determinant $\Delta_n(f, b, \sigma)$ of the coefficient matrix vanishes. Since the full form of the determinant is rather unwieldy, we do not present it here, but give the full expression in the appendix \ref{app1}. We note that for a given coefficient $b$ and fixed elasticity properties, the determinant is a function of the dimensionless frequency $f$ only, since $x_{\pm}$ are themselves functions of $f$. On general grounds, the frequencies $f$ which yield a vanishing determinant will be at most a discrete set of values. The reason for this is that as the frequency $f$ is varied, the determinant, being a combination of Bessel functions and derivatives, will oscillate and change its sign. Each zero of the determinant is a normal frequency of elastic vibrations. Below we carry out numerical calculations to determine some of these frequencies.

Solving the equation $\Delta_n(f, b, \sigma)=0$ for $f$ gives a frequency spectrum $f(n,s,\sigma,b)$, with the integer $s\geq 0$ numbering the roots for given $n,\sigma,b$. One property of these solutions is that they are invariant under the simultaneous change of signs $f \rightarrow -f, b \rightarrow -b$. This is a consequence of the fact that our elastic system is time-reversal invariant if the magnetic field is simultaneously flipped as well. However, unlike in the zero field case, the solutions are no longer invariant under $n \rightarrow -n$ only, because the Lorentz term breaks the time-reversal symmetry of the problem.

\section{Spectrum of oscillations} \label{sec:oscSpec}
We now explore the solutions of the characteristic equation \eqref{eq:master} and how they depend on the magnetic field $b$ and the angular wavenumber $n$. We find that the normal modes with $s\geq 1$ do not change qualitatively for different values of the Poisson ratio $\sigma$. For this reason we fix the latter to a representative value of $\sigma = 0.25$ throughout this section. On the other hand, we explain in sec. \ref{sec:NewSurfModes} that the lowest normal modes with $s=0$ correspond precisely to the Rayleigh-waves. Their spectra have a strong dependence on the Poisson ratio.

\begin{figure*}[t!]
\begin{subfigure}{.4\textwidth}	
	\includegraphics[width=\textwidth ]{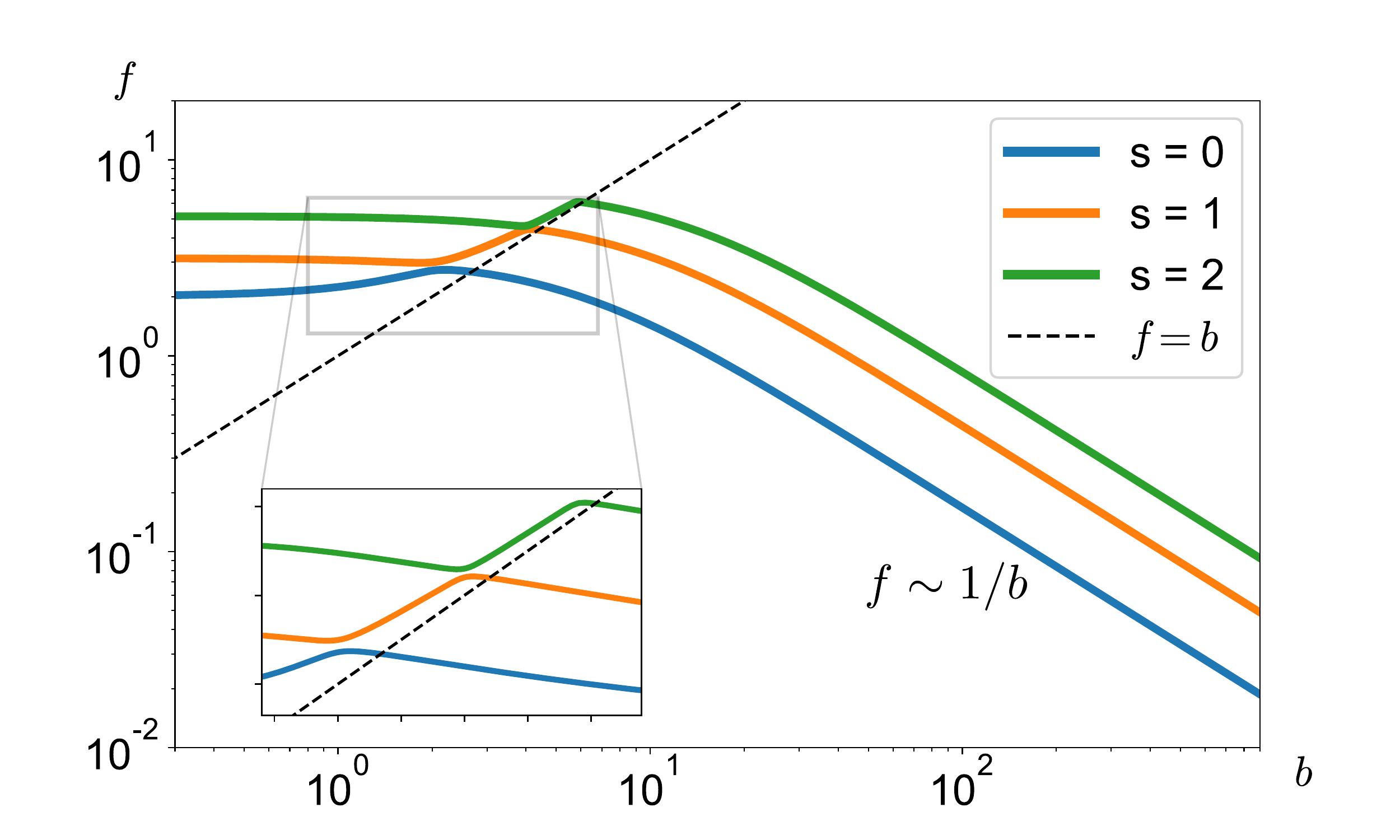}
	\caption{}
	\label{fig:f(b)_n0}
\end{subfigure}%
\begin{subfigure}{.4\textwidth}	
	\includegraphics[width=\textwidth ]{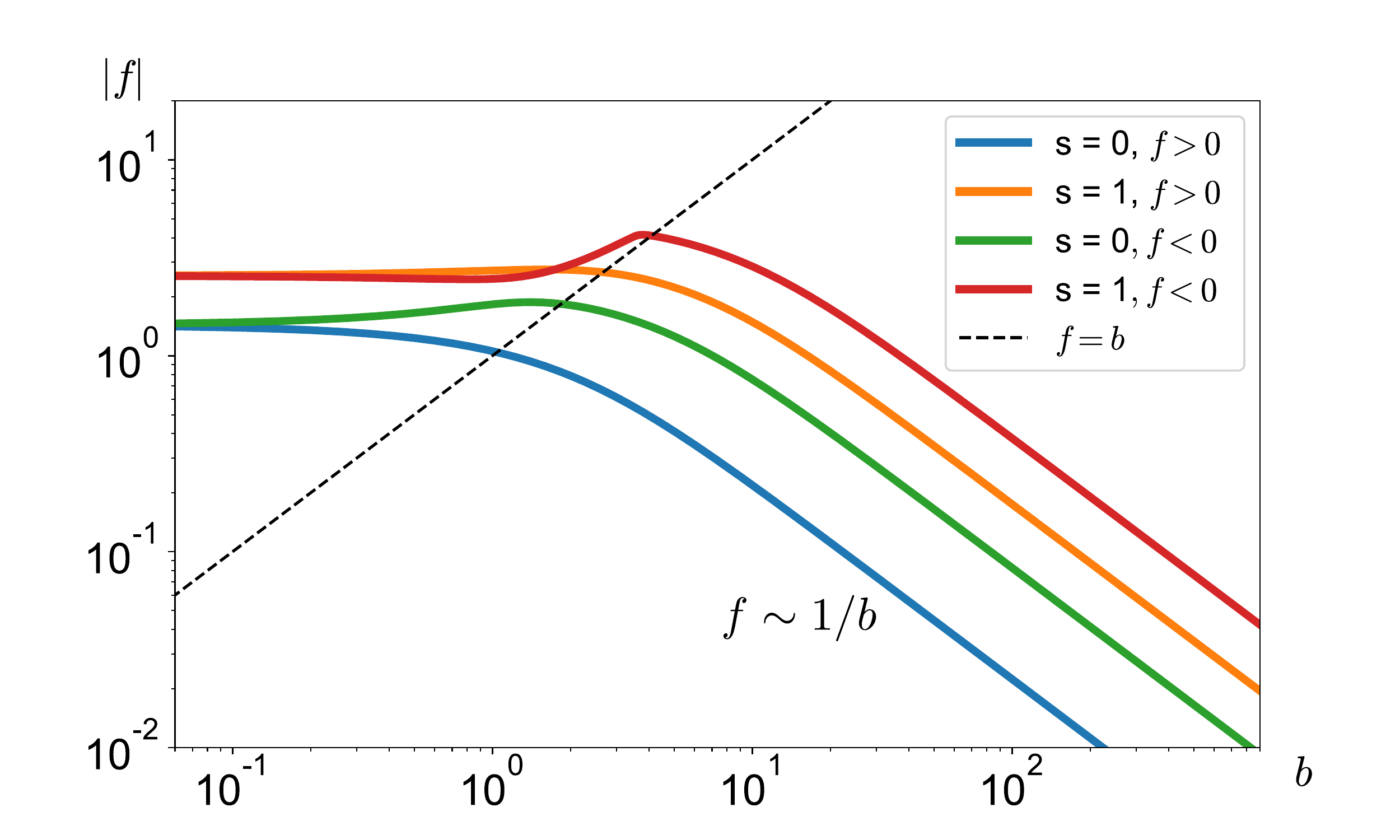}
	\caption{}
	\label{fig:f(b)_n2}
\end{subfigure}
\caption{(a) Frequency dependence on the dimensionless parameter $b$ with $n=0$,  $\sigma = 0.25$ for the three lowest modes $s=0,1,2$. The inset shows the avoided crossings near the line $f=b$. (b) Frequency dependence on the dimensionless parameter $b$ with $n=2$,  $\sigma = 0.25$ for the lowest two modes $s=0,1$. Both the negative and positive frequency solutions are shown.}	
\end{figure*}
\begin{figure*}[t!]
\begin{subfigure}{.4\textwidth}	
	\includegraphics[width=\textwidth ]{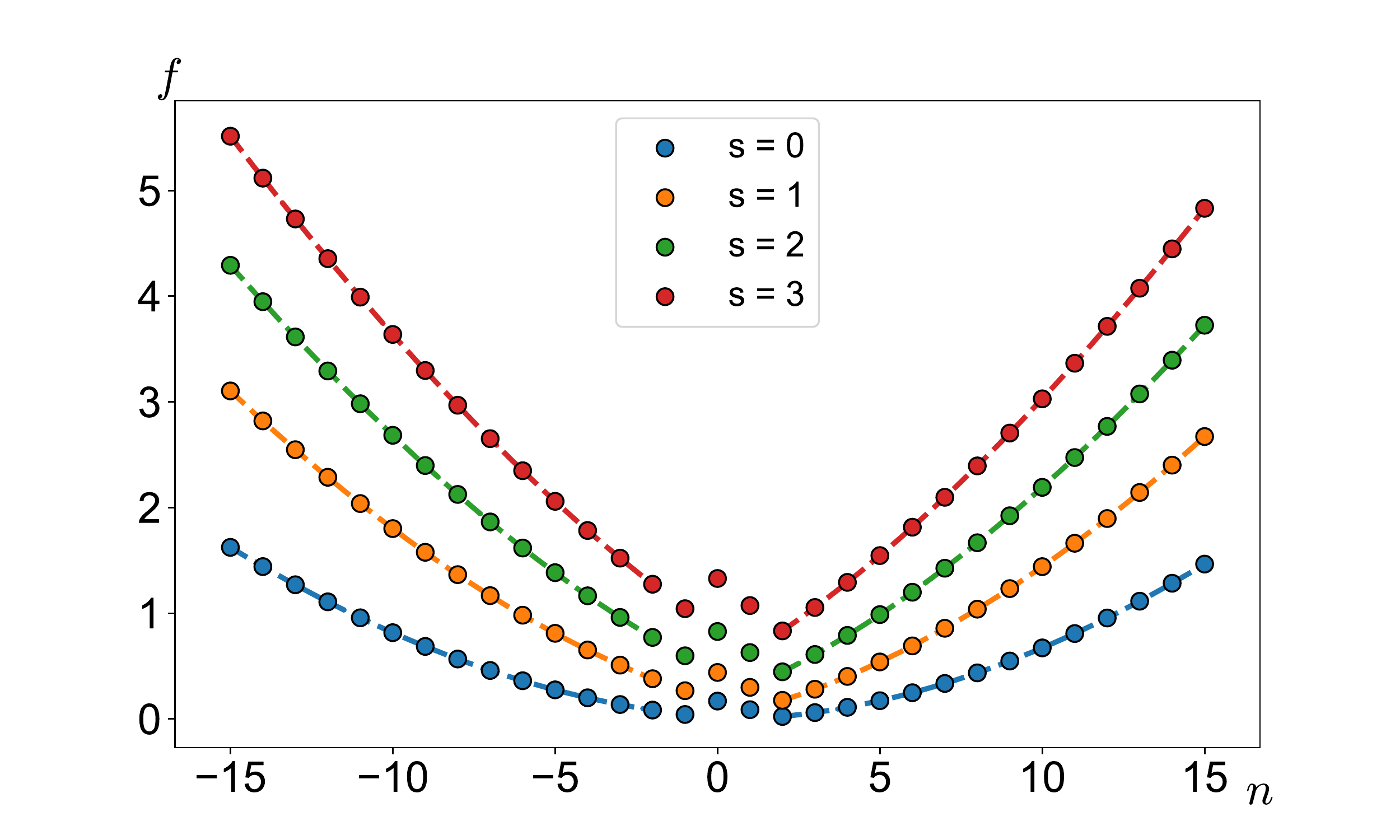}
	\caption{}
	\label{fig:f(n)_sigma_25_highb}
\end{subfigure}%
\begin{subfigure}{.4\textwidth}	
	\includegraphics[width=\textwidth ]{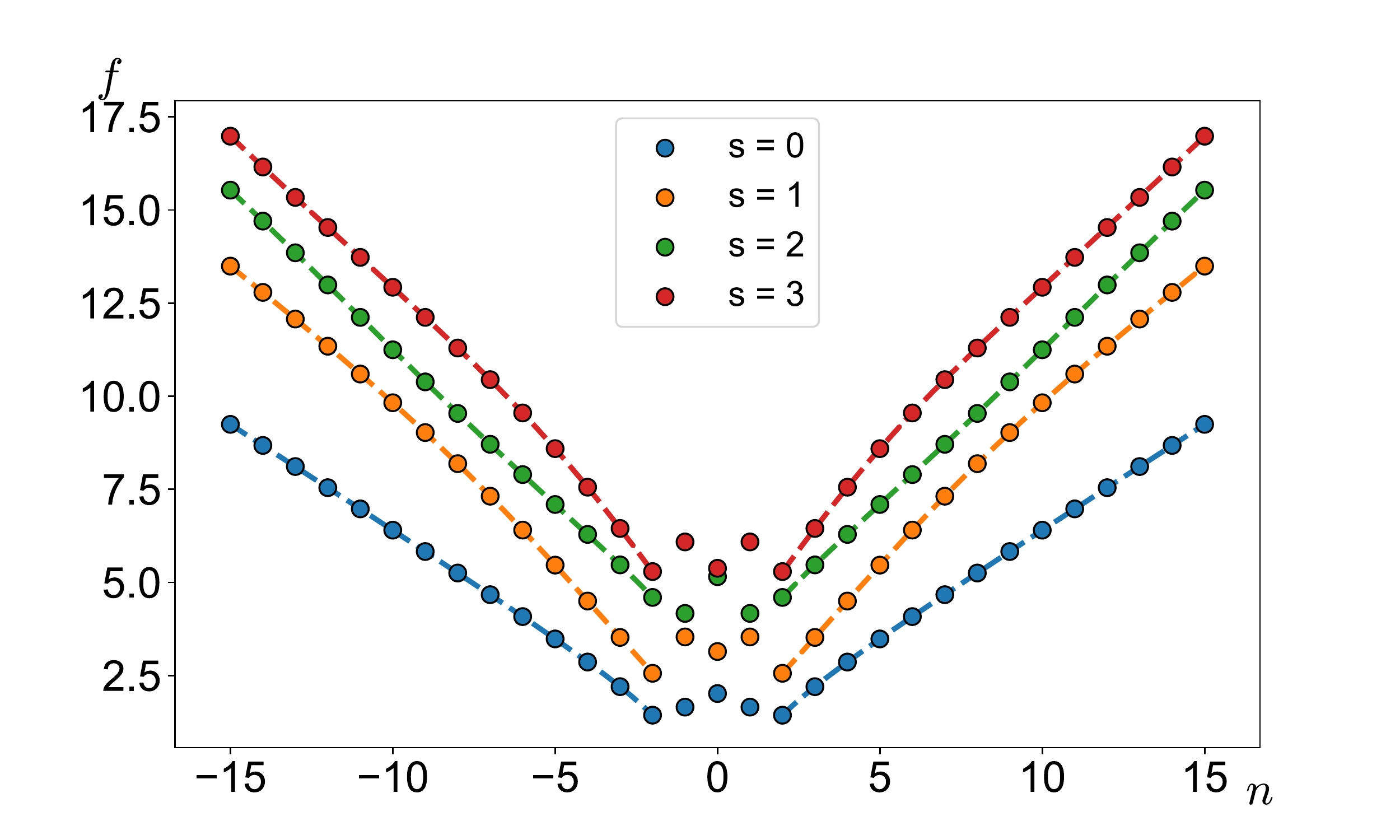}
	\caption{}
	\label{fig:f(n)_sigma_25_lowb}
\end{subfigure}
\caption{Frequency dependence on the angular wavenumber $n$ with (a) $b=100$ and (b) $b=0$, $\sigma = 0.25$ for the lowest four modes. Dashed lines join the points to aid the eye.}	
\end{figure*}

\subsection{Frequency dependence on magnetic field}
The complicated form of the characteristic equation \eqref{eq:master} requires us to treat the full problem numerically. The exact numerical solution of the characteristic equation yields the eigenfrequencies as a function of the magnetic field. The plots in Figs. \ref{fig:f(b)_n0}, \ref{fig:f(b)_n2} show this dependence for $n=0$ and $n=2$, respectively.

Numerically one finds that for small fields $b$ the frequency grows linearly with $b$. 
It turns out that in the high field regime $|b| \gg 1$ the characteristic equation becomes analytically tractable, providing us with an understanding of this limit.
In particular, for the high-field regime the dispersion \eqref{eq:disp} reduces to
        \begin{equation}
        	|f| = \pm \frac{1}{b} \sqrt{\frac{1-\sigma}{2}} x^2_\pm.
        	\label{eq:disp_red}
        \end{equation}
Inserting this into the characteristic equation \eqref{eq:master}, we find that $x_\pm$ becomes independent of $b$ in the large $b$ limit and from eq. \eqref{eq:disp_red} we deduce a $1/b$ behavior of the frequency. To summarize we have the two asymptotic relations
\begin{equation}
f \sim
\begin{cases}
c_0 + c_1 b& \text{as } b \rightarrow 0 \\
1/b, & \text{as } b \rightarrow \infty .
\end{cases}
\label{eq:f_asymp}
\end{equation}
with $c_0$ and $c_1$ parameters that depend on $(\sigma,n,s)$. We checked that the function $c_0(\sigma, n, s)$ is in agreement with results previously obtained by \cite{Holland1966, Bashmal2010} when the field $b$ is zero.
\begin{figure*}[t!]
\begin{subfigure}{.4\textwidth}	
	\includegraphics[width=\textwidth ]{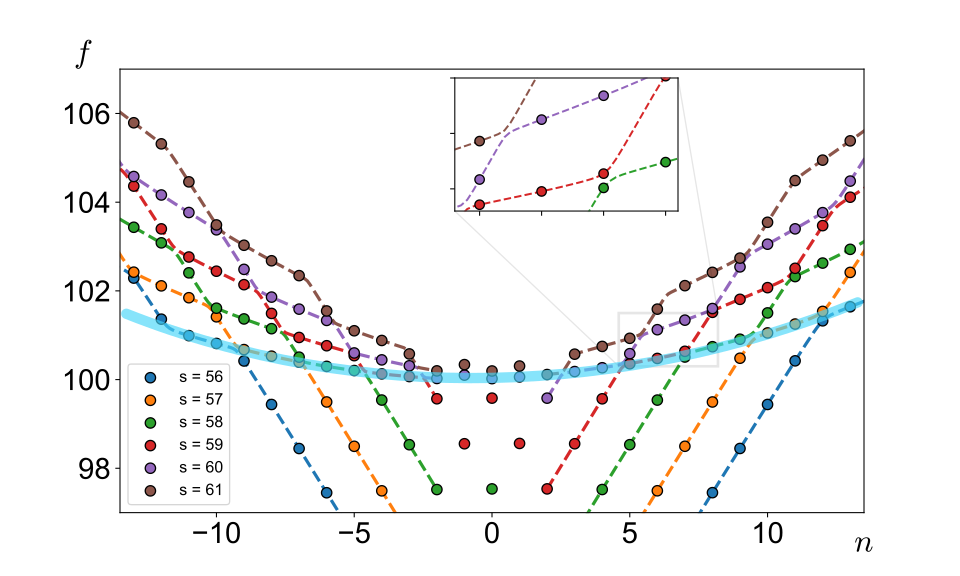}
	\caption{}
	\label{fig:f(n)_Kohn_sigma_25_quad}
\end{subfigure}%
\begin{subfigure}{.4\textwidth}	
	\includegraphics[width=\textwidth ]{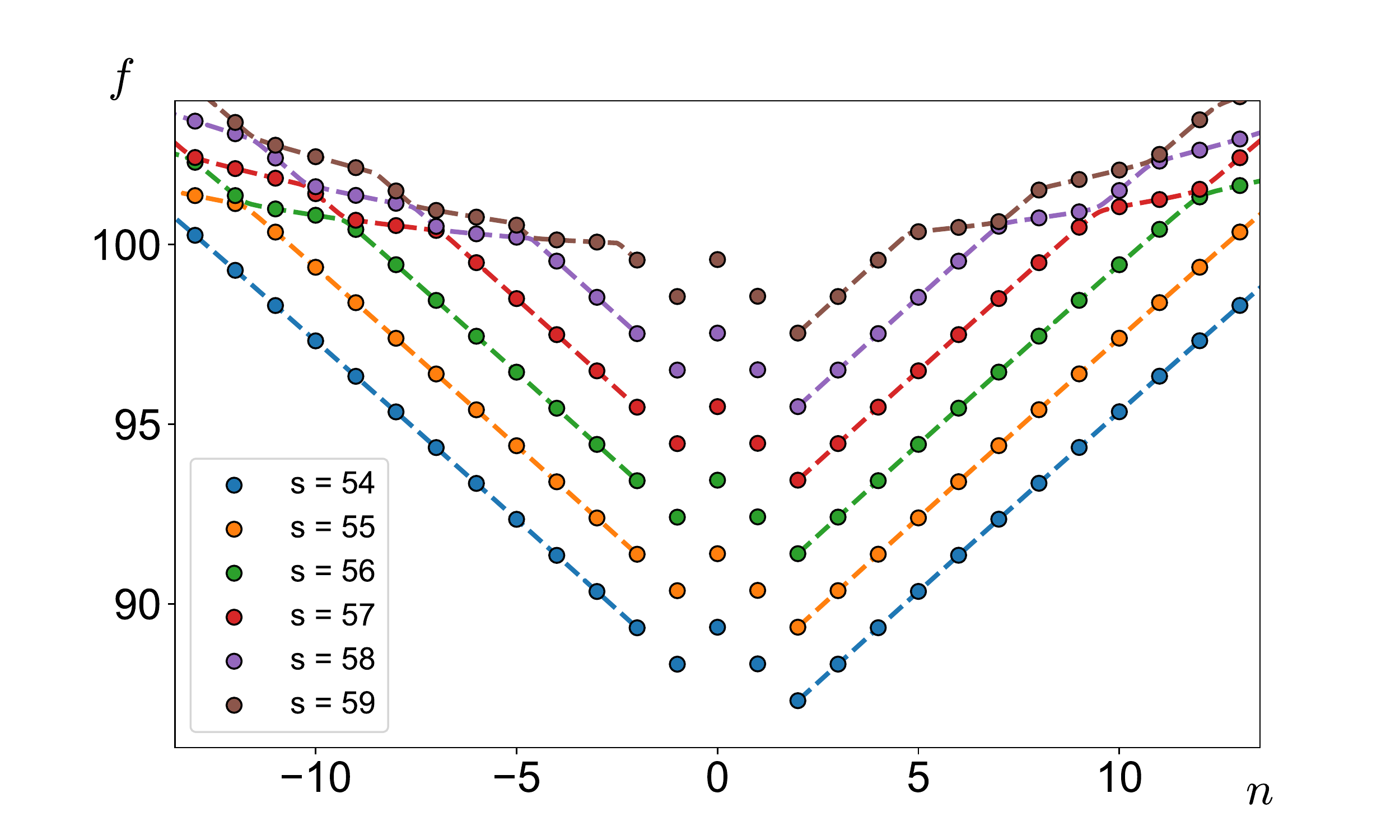}
	\caption{}
	\label{fig:f(n)_Kohn_sigma_25_linear}
\end{subfigure}
\caption{(a) Frequency dependence on the angular wavenumber $n$ with $b=100$, $\sigma = 0.25$, plotted for modes with (a) $f\gtrsim b$ and (b) $f\lesssim b$, respectively. Dashed lines join the points to aid the eye. We have marked the branch with cyclotron surface modes in turquoise color.}
\end{figure*}
Interestingly, we observe that for intermediate values of $b$ the frequency curves with different values of $s$ repel each other, see the inset of Fig. \ref{fig:f(b)_n0}. A discussion of this phenomenon is given at the end of the next subsection.
								
As noted above, the Lorentz term breaks the time reversal symmetry. As a consequence the spectrum is not invariant under $f\rightarrow -f$. The breaking of this symmetry with increasing value of $b$ is demonstrated in Fig. \ref{fig:f(b)_n2}, where the negative and positive eigenfrequencies are shown for the $n=2$ case. An implication of this time-reversal breaking is that we cannot form superpositions of time-reversed eigenmodes that produce standing waves, see the discussion in sec. \ref{sec:Shape_oscillations}.

\subsection{Frequency dependence on angular wavenumber}
\label{sec:FreqDisp}

We now turn our attention to the study of the dispersion of the frequency spectrum with the angular wavenumber $n$. For fixed integers $n$, the characteristic equation has as solutions a discrete sequence of eigenfrequencies. 
The quadratic dependence on the angular wavenumber can clearly be observed in Fig. \ref{fig:f(n)_sigma_25_highb}, where we show the dispersion for a high field value of $b=100$. For comparison Fig. \ref{fig:f(n)_sigma_25_lowb} shows the $b=0$ result. Note that in order to aid the eye, we are displaying dashed lines that connect the discrete sequence of eigenfrequencies. These lines are generated by solving the characteristic equation at non-integral values of $n$. These values are unphysical, but continuing $n$ to the real line is a mathematically well-defined procedure and can therefore be used to aid in the visualization.

The high field shape of the spectrum is quite different from the zero field case. When the field $b$ is zero, the eigenfrequencies grow linearly with $n$, for large $|n|$ \cite{sezawa1927dispersion}. When the magnetic field is finite and the considered frequencies are less than $b$, the dispersion relation changes from linear to quadratic. We call this range of frequencies the Lorentz-dominated regime.   Furthermore, the field also breaks the $n \rightarrow -n$ symmetry of the spectrum. 
\newline

Interestingly, the dispersion shows new features when one zooms in on the frequency region near $f=b$, see Figs. \ref{fig:f(n)_Kohn_sigma_25_quad} and \ref{fig:f(n)_Kohn_sigma_25_linear}. Quite strikingly one observes that the frequency values lie on a network formed out of parallel-translated parabolas and lines. In the following we give an analytic explanation of this phenomenon. 
We focus on the $f \approx b$ regime with $b\gg 1$. For $f \lesssim b$, in eq. \eqref{eq:disp} $x_-$ is imaginary and thus in eqs. \eqref{eq:solnOrder1} and \eqref{eq:solnOrder2} only the $q_+$ part oscillates. Upon crossing $f=b$, both parts oscillate, here $x_-$ is very small while $x_+$ is very large.
We can analytically investigate the regime of interest by writing $f = b+ \Delta f$ with a small $\Delta f$ and expanding the dispersion relation \eqref{eq:disp} to linear order in $\Delta f$. Inverting eq. \eqref{eq:disp}, yields the two families of frequencies
\begin{eqnarray}
\Delta f  &=& - \frac{b}{ 1 - \frac{2(1-\sigma)}{(3-\sigma)^2} } + \frac{\sqrt{\frac{1-\sigma}{3-\sigma}}}{ 1 - \frac{2(1-\sigma)}{(3-\sigma)^2} }x_+  \label{eq:approximated_x+}  \\
\Delta f &=& \frac{3-\sigma}{4b} x^2_- .
\label{eq:approximated_x-}
\end{eqnarray}
We show in the appendix \ref{app:prooflinear} that for large $n$, both $x_+$ and $x_-$ grow linearly with $n$. As a consequence, the linearly spaced $x_+$ solutions yield, according to eq. \eqref{eq:approximated_x+}, the family of linearly spaced frequencies. While the linearly spaced values of $x_-$ yield according to eq. \eqref{eq:approximated_x-} the family of quadratically spaced frequencies. This explains the network of parabolas and lines found in the Figs. \ref{fig:f(n)_Kohn_sigma_25_quad} and \ref{fig:f(n)_Kohn_sigma_25_linear}.
The lowest parabolic series of points, highlighted by the turquoise colored curve in Fig. \ref{fig:f(n)_Kohn_sigma_25_quad}, turns out to be a special kind of surface wave that is distinct from Rayleigh waves. We refer to it below as the cyclotron surface wave, since it has a frequency close to the cyclotron frequency. We will discuss this surface wave in some detail in sec. \ref{sec:NewSurfModes}.

Another interesting feature of these plots is the fact that the dashed curves never cross each other. 
This aspect can be understood intuitively by examining the form of the matrix equation \eqref{eq:det}. Let us consider two dashed curves with consecutive values of $s$. By definition they are solutions to the characteristic equation $\Delta_n(f(s), b, \sigma) = 0$ and $\Delta_n(f(s+1), b, \sigma) = 0$ respectively, with $n$ viewed as a continuous variable. A crossing point is unlikely, since it would imply that $\Delta_n(f,b,\sigma)$, which is a complicated function of Bessel functions, must have a double root at a particular value of $n$. 
We note that the avoided crossings in Figure \ref{fig:f(b)_n0} have the same mathematical origin, except that here the double root of $\Delta_n(f,b,\sigma)$ would occur as a function of the parameter $b$.

\section{Surface modes-- Rayleigh and beyond} \label{sec:NewSurfModes}
\subsection{Rayleigh modes}
In semi-infinite elastic systems a particular type of mode can be excited that is exponentially localized near the edge of the material, the so-called Rayleigh wave \cite{Rayleigh1885, landauElasticity, thorne}. In the case of a finite disk, however, there is no sharp distinction between propagation on the edge and propagation in the bulk \cite{sezawa1927dispersion}. 
As clarified by Viktorov \cite{viktorov1967rayleigh}, for large enough $n$ it is the lowest $s=0$ branch which becomes the Rayleigh mode as the disk radius $R$ tends to infinity. In App. \ref{app3} we sketch his argument and show that it also applies in the presence of the Lorentz term. 

In the previous work \cite{2021rayleigh} it was found that crystals with the Lorentz term support Rayleigh waves in the half-plane geometry for all values of the Poisson ratio $\sigma$. It was also demonstrated that the wave propagation is asymmetric with a dispersion relation that is dependent on the sign of the wavenumber $k$. The notable exception is the special point $\sigma = 1/3$, where the Rayleigh wave spectrum becomes symmetric. Naturally, in the disk geometry for large enough $|n|$, we must recover all previous results, since waves with a large azimuthal wavenumber $|n|$ have short wavelengths and are therefore insensitive to the curvature of the edge.  However, for small $|n|$ departures from the semi-infinite results are to be expected. 

To capture the effect of edge curvature, we have studied the Rayleigh wave spectrum for $\sigma=1/3$ and finite values of $b$. \begin{figure}
	\includegraphics[width=0.8\columnwidth ]{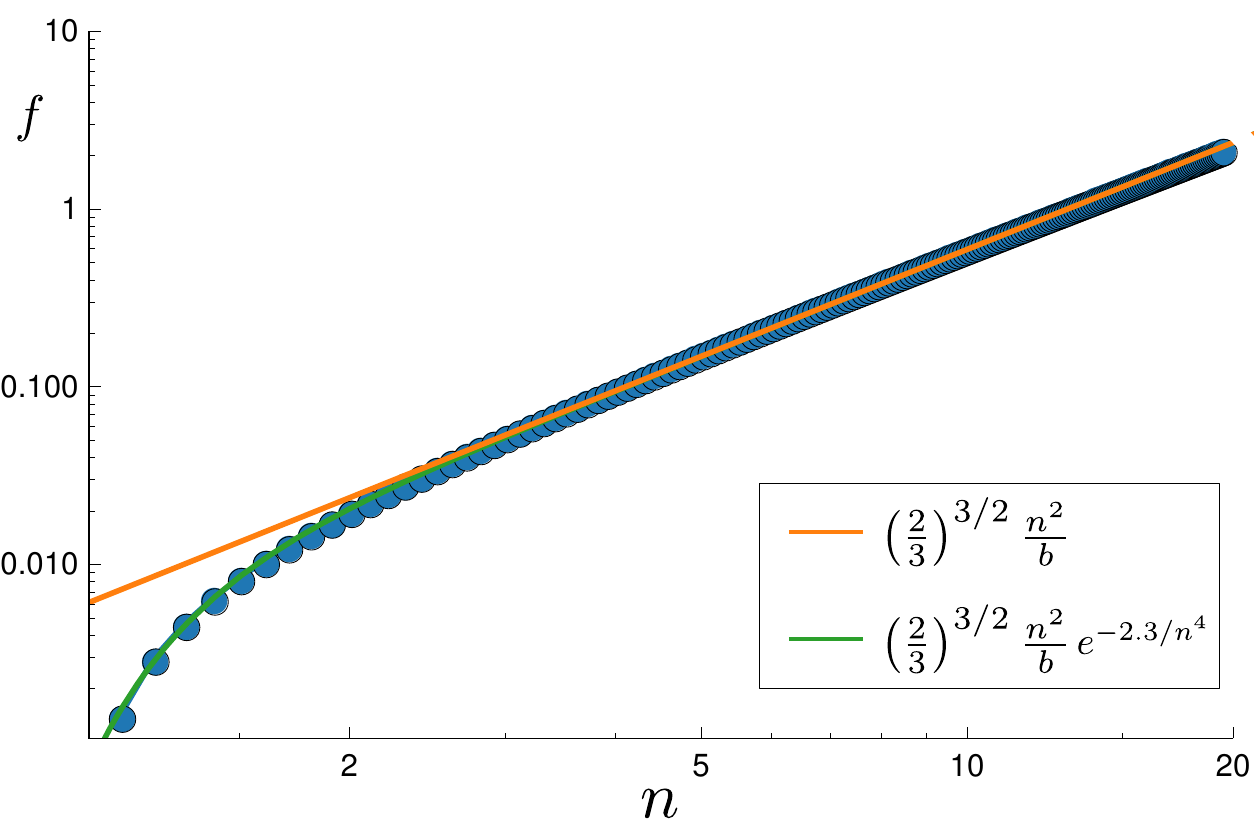}
	\caption{Frequency $f$ of the Rayleigh modes ($s=0$) as a function of the angular wavenumber $n$ for $b=100$ and $\sigma=1/3$. At large $n$ the frequency scales quadratically in $n$ according to our result in \cite{2021rayleigh}. The frequency is suppressed at small values of $n$.}
	\label{fig:LogLogFitDisp}
\end{figure}
Shown as blue circles in Fig. \ref{fig:LogLogFitDisp} are the frequencies $f$ of the $s=0$ normal modes as a function of the angular wavenumber $n$ for $b=100$ in a log-log plot. When $n$ is large, such that we are in the Lorentz-dominated regime, the dispersion follows a quadratic behavior (orange line), with a prefactor that exactly matches the analytical result obtained in \cite{2021rayleigh}. At smaller $n$, however, the frequencies are substantially suppressed.  A fit (green) to the data with exponentials of inverse powers of $n$ shows that a dispersion of the form
\begin{eqnarray}
f(k_\varphi) = \left(\frac{2}{3}\right)^{3/2}\frac{k_\varphi^2 R^2}{b} \times e^{-2.3/k_\varphi^4 R^4}
\label{eq:RayleighCrossover}
\end{eqnarray}
captures the crossover well. The first factor is our analytical result \cite{2021rayleigh}. Interestingly, we find that the exponential factor is identical for different values of $b$. Here we have explicitly used the azimuthal wavenumber $k_\varphi = n/R$ in order to emphasize the dependence on the disk radius. The lower value of the frequency for the disk compared to the semi-infinite medium stems from the low-frequency shape-oscillations discussed in the next section.   For $R\rightarrow \infty$ the exponential factor tends to $1$ and we recover the scaling of the infinite system.

\subsection{Shape oscillations}
\label{sec:Shape_oscillations}
\begin{figure*}
	\includegraphics[width=0.65\textwidth]{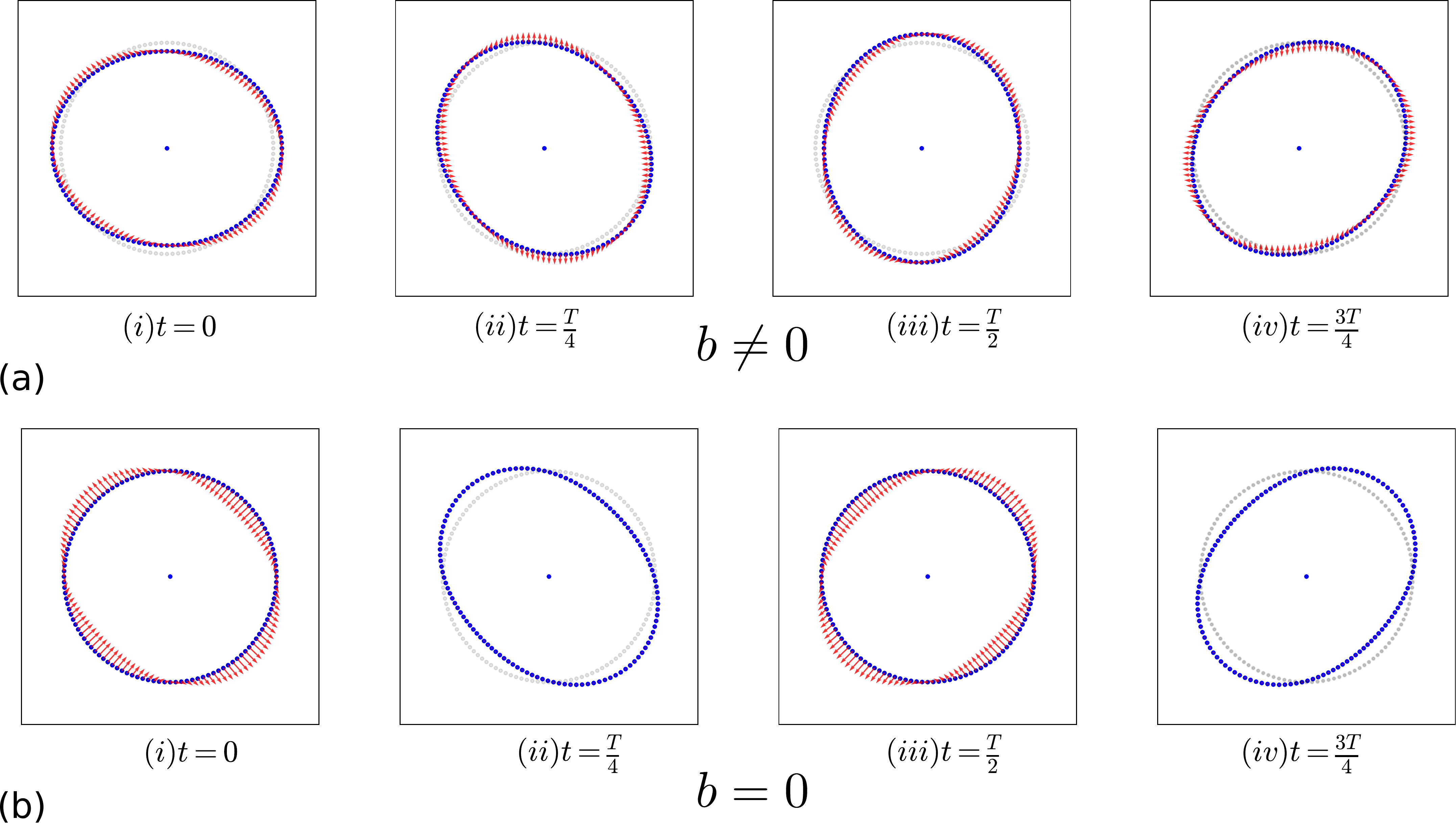}
\caption{Time series of shape oscillations, for videos see \cite{supmat}. (a) Oscillation pattern for $n=2$, $b=1000$, $\sigma = 0.25$ and $s=0$. The blue points show the disk boundary displaced from the equilibrium position (gray). The velocity vector-field is shown in red. During one period of oscillation the pattern rotates as a whole by an angle $2\pi$.  (b)  When the field is turned off for the same set of parameters the oscillation pattern becomes a standing wave.}	 
\label{fig:shapeosc}
\end{figure*}

A property that is of particular interest to experimental observations are the low-frequency shape oscillations of the elastic disk. To study these, we turn to the form of the eigenmodes given in eqs. \eqref{eq:displace} and \eqref{eq:displacetheta} of the appendix \ref{app2}. The numerical solutions for the $n=2$, $s=0$ eigenmode in the presence and in the absence of the magnetic field, respectively, are visualized in videos \cite{supmat} and Fig. \ref{fig:shapeosc}. The oscillation pattern of this figure shows two maxima. In general the oscillation with angular wavenumber $n$ has $n$ maxima, since the potentials have an angular dependence given by $\exp(i n \theta)$. 
\begin{figure}[t!]
\includegraphics[width=\columnwidth]{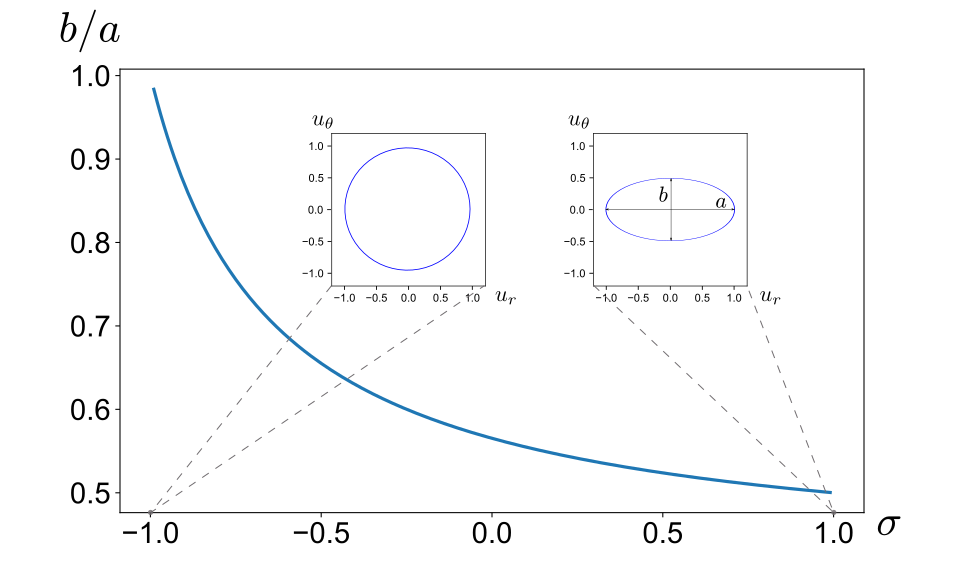}
\caption{The trajectory of a point on the boundary during a full period of oscillation is elliptical. The aspect ratio $b/a$ is shown as a function of the Poisson ratio $\sigma$. Plotted are the radial versus the azimuthal displacements for $b = 100, n = 2, s = 0$.}
\label{fig:pointosc}
\end{figure}
The shape of the disk outline always rotates as a whole. The direction can be flipped by switching the sign of the magnetic field $b$. However, in the zero-field case, a standing field oscillation pattern is always an eigenmode, since one can superpose an oscillation with frequency $\omega$ with its time-reversed partner, which has frequency $-\omega$.  Such a superposition is not an eigenmode for finite $b$, since broken time-reversal implies that the time-reversed partner is not a solution of the equations of motion.

The shape oscillations shown in Fig. \ref{fig:shapeosc} are not qualitatively affected by the value of the Poisson ratio $\sigma$. 
However, as $\sigma$ is tuned, the polarization of the oscillation changes. We follow this change by tracing out the trajectory of a point on the boundary during a single period of oscillation for the $n=2, s=0$ mode. We find that the shape is always an ellipse. As the value of the Poisson ratio is tuned from $-1$ to $+1$, we find that the trajectory starts out as a circle and is continuously deformed into ellipses of increasing eccentricity, see Fig. \ref{fig:pointosc}. This is in agreement with the finding in \cite{2021rayleigh} that at $\sigma = -1$ the boundary hosts a special mode with frequency $f=b$ that is circularly polarized.

\subsection{Cyclotron Surface Mode}
\begin{figure*}[t]
\includegraphics[width=0.7\textwidth]{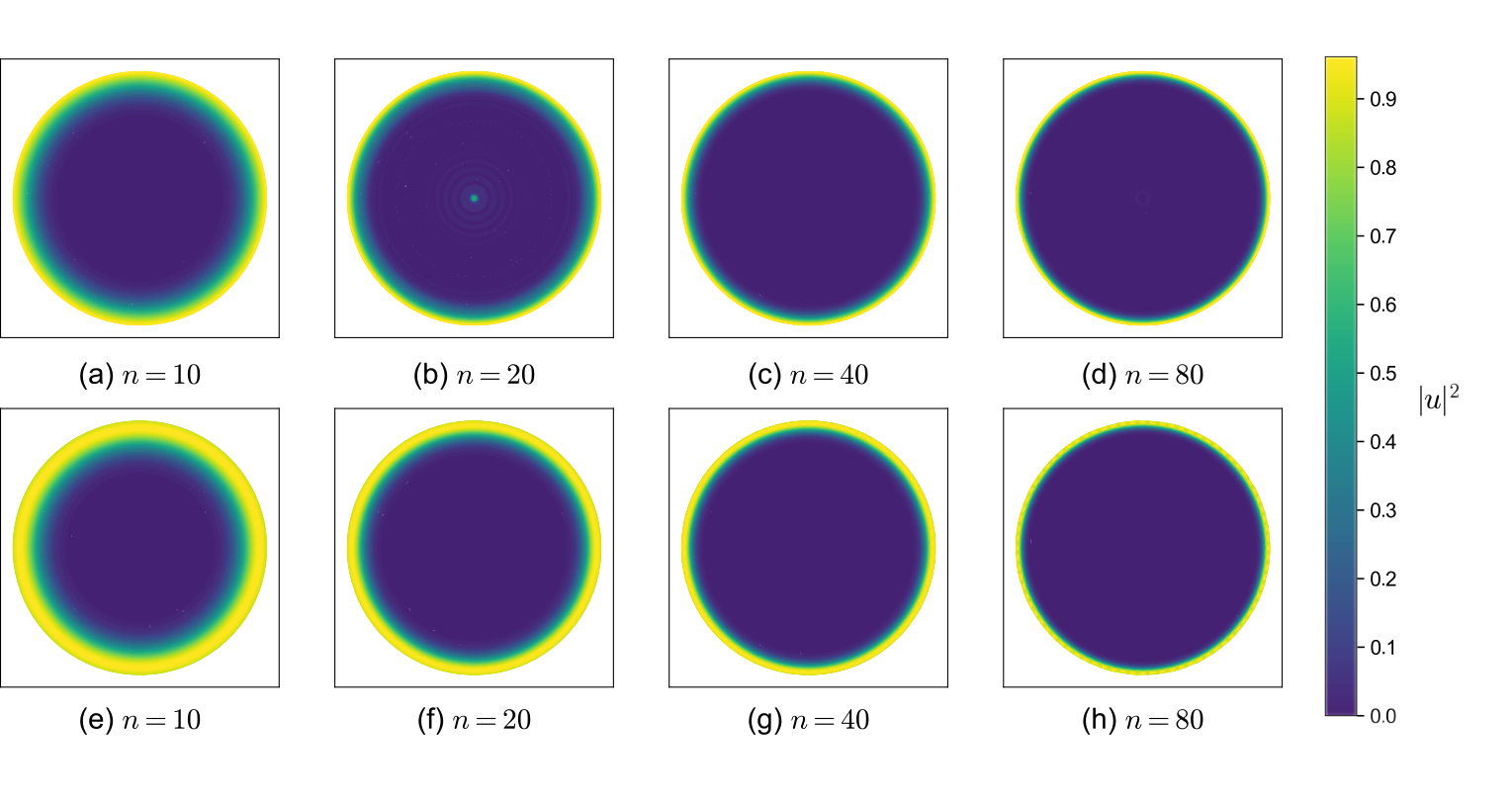}
\caption{\label{fig:surface} Plot of normalized time averaged kinetic energy density throughout the disk of radius $R$ for $b=1000$, $\sigma = 0.25$. (a-d) Cyclotron surface waves for different values of $n$. (e-h) Rayleigh-type $s=0$ modes for the same values of $n$.\newline
Both types of waves show a strong concentration of kinetic energy near the boundary. For the cyclotron surface waves there is, in addition, a small excited region near the center of the disk. The displacement amplitude is modulated with wavenumber $q_\text{pattern}$, see eq. \eqref{eq:qpattern}.} 
\end{figure*}	
Now we turn to the study of the cyclotron surface mode that we mentioned in sec. \ref{sec:FreqDisp} and that we highlighted in turquoise color in Fig. \ref{fig:f(n)_Kohn_sigma_25_quad}. In Figs.  \ref{fig:surface} (a-d) these modes are visualized by density plots of the time-averaged kinetic energy of the disk. Clearly, most of the kinetic energy is localized near the boundary of the disk, indicating that these are well-localized surface modes.  As a reference, we are also plotting the Rayleigh modes ($s=0$) in the same way in the lower row (e-h). The degree of localization is similar in both cases. However, notice that some of the cyclotron surface modes, as for example the one in (b), have a noticeable kinetic energy density very close to the center of the disk.

As was discussed in sec. \ref{sec:FreqDisp} in the region $f\approx b$, where the cyclotron surface mode appears, the value of $x_-$ is small compared to unity, while the value of $x_+$ is very large. The full displacement field is a superposition of the $+$ and $-$ branches as given in Appendix \ref{app2}, see eqs. \eqref{eq:displace} and \eqref{eq:displacetheta}. As a consequence of the smallness of $x_-$, the $-$ branch contribution to the displacement field is nearly vanishing close to the center and monotonically increases towards the edge. If $n \gtrsim 10$ the $-$ branch is well-localized near the edge. A similar statement also holds for Rayleigh waves in a disk geometry \cite{sezawa1927dispersion}. If we can now identify an interval of $n$ values for which the + branch contribution in eqs. \eqref{eq:displace} and \eqref{eq:displacetheta} remains small, then the resulting modes must be essentially edge-modes, since they will be well-localized near the disk boundary. 

In the appendix \ref{app2} we show that for $x_+ \gg n$ the contributions from the $+$ branch are negligible, see eqs. \eqref{eq:asymptoticsContribr} and \eqref{eq:asymptoticsContribTheta}. 
Numerically, we identify that this inequality holds whenever $n < b/10$. To summarize, inside the window $10 < n < b/10$ we find well-localized surface modes that we named cyclotron surface waves and which are distinct from Rayleigh waves.

We derive the explicit expression for the cyclotron surface eigenmode in the appendix, see eqs. \eqref{eq:newmodes_r} and \eqref{eq:newmodes_t}. 

The question now arises as to how these modes can be distinguished in experiments from the quadratically dispersing Rayleigh waves with $s=0$. First, in contrast to the latter, the frequency-scale of the cyclotron surface waves is set by the frequency $\Omega$, while the Rayleigh mode frequencies are much smaller. Moreover, in the case of the cyclotron surface modes there is a small region at the center of the disk that is noticeably displaced and thereby serves as a spatial signature. Its origin stems from a small contribution of the $+$ mode as discussed above. In fact, for $n=20$ this effect is quite strong, and can be seen as a series of concentric rings in the displacement field, see Fig. \ref{fig:surface}(b). We recall that the presence of this small $+$ branch contribution is in the end a consequence of the fact that a true eigenmode has to satisfy the stress-free boundary conditions \eqref{eq:boundary1} and \eqref{eq:boundary2}. 
	
We can work out the shape of this signature pattern by expanding eqs. \eqref{eq:displace} and \eqref{eq:displacetheta} for large $x_+ \gg 1$ and dropping the terms involving $x_-$. In this regime we can employ the large-argument asymptotics for $J_n(x)$ \cite{Watson1995} to derive the wavenumber of the oscillation, we find
\begin{equation}
q_\text{pattern} \approx q_+ \approx \left[\frac{1}{2 v_1+v_2}+\frac{1}{v_2}\right]^{1/2} \Omega.
\label{eq:qpattern}
\end{equation}
Clearly this wavenumber is entirely determined by the elastic properties of the material and the cyclotron frequency.
A small-amplitude pattern at the center of the disk with the characteristic wavenumber $q_\text{pattern}$ is the hallmark of this excitation that is almost fully localized on the edge. 

\section{Conclusion and Outlook}
\label{sec:concl}
In this paper we explored the normal modes of an elastic disk described using the effective field theory \eqref{eq:Langrangian} with a Lorentz term that breaks time reversal and parity symmetries. By contrasting the spectrum with the results that some of us previously obtained for the semi-infinite plane \cite{2021rayleigh}, we uncover the long-wavelength modifications for the disk geometry. The most striking of our results is the presence of strongly localized cyclotron surface modes, which are different from the classical Rayleigh waves. We expect that these modes are experimentally observable in networks of coupled gyroscopes. We predict using the elasticity theory \eqref{eq:Langrangian} that the cyclotron surface modes can be recognized by the presence of a series of very weak rings near the center of the disk with a modulation wavenumber given in eq. \eqref{eq:qpattern}. 

The fact that the cyclotron surface mode appears at frequencies close to the cyclotron frequency $ \Omega$ implies that higher order terms in the effective Lagrangian \eqref{eq:Langrangian}, which were not considered in the present work, may play an important role. It will therefore be of considerable interest to probe the robustness of our predictions by performing surface wave experiments on gyroscopic metamaterials at frequencies close to the cyclotron frequency $\Omega$.

\begin{acknowledgements}
We would like to acknowledge useful discussions with Egor Kiselev.
Our work is funded by the Deutsche Forschungsgemeinschaft (DFG, German Research Foundation) under Emmy Noether Programme grant no.~MO 3013/1-1 and under Germany's Excellence Strategy - EXC-2111 - 390814868.
\end{acknowledgements}

\appendix
\begin{widetext}
\newpage
\section{The characteristic equation}
\label{app1}
The characteristic equation for the eigenfrequencies of the vibrational modes of the disk is obtained by setting the determinant of the matrix in eq. \eqref{eq:det} equal to zero. The polynomials appearing in that equation are given by
\begin{eqnarray}
P_n(x,\alpha) &=& J_n(x) [ n^2 - x^2 + n\alpha - n^2 \sigma - n\alpha \sigma]  + x J'_n(x) [ -1 - n\alpha + n\alpha \sigma + \sigma ], 
\label{eq:poly1}
\\
Q_n(x,\alpha) &=& J_n(x) [-2 n^2\alpha - 2n + \alpha x^2 ] + 2x J'_n(x) [n + \alpha ],
\label{eq:poly2}
\end{eqnarray}			
where $\alpha^\pm$ is the ratio of the amplitudes $B^\pm_n / A^\pm_n$ obtained from eq. \eqref{eq:matrix}
\begin{equation}
\alpha^\pm = -\frac{f}{2b} \cdot \frac{1+\sigma}{1-\sigma} \left[1 \pm \sqrt{1+ 8 \frac{b^2}{f^2} \cdot \frac{1-\sigma}{(1+\sigma)^2}}~\right].
\label{eq:ratio}
\end{equation}
With these results we can write the characteristic equation in the final form
\begin{equation}
\begin{split}
&\  J_n(x_+)J_n(x_-) \cdot 2f^2 \sqrt{\chi} ~ (1+\sigma) \cdot 
 \left[ -\frac{f^3}{b}\cdot \frac{1}{(1-\sigma)^2} + f \left( \frac{2n(n-1)}{(1-\sigma)b} + \frac{b}{(1-\sigma)^2}\right) -\frac{2n(n-1)}{(1-\sigma)}\right] \\
&\ + J_n(x_-)J_{n+1}(x_+)x_+ 
 \left[ -\frac{f^3}{b} (1+\sigma)\cdot \right.\frac{3-\sigma - (1+\sigma)\sqrt{\chi}}{4(1-\sigma)} \cdot 
 \left( 1 - \sqrt{\chi} - \frac{2}{1-\sigma}(1+\sqrt{\chi}~)\right) \\
 &\ - 2nf^2 \cdot \frac{3-\sigma - (1+\sigma)\sqrt{\chi}}{1-\sigma} 
 \left.- \frac{f}{b}\sqrt{\chi}(1+\sigma)2n(n-1)(n+1)  + 2n(n-1)(n+1)(3-\sigma) \right]\\
 &\ - J_n(x_+)J_{n+1}(x_-)x_- \left[ -\frac{f^3}{b} (1+\sigma)\cdot \frac{3-\sigma + (1+\sigma)\sqrt{\chi}}{4(1-\sigma)} \left( 1 + \sqrt{\chi} - \frac{2}{1-\sigma}(1-\sqrt{\chi})\right) \right.\\
&\	\left. - 2nf^2 \cdot \frac{3-\sigma + (1+\sigma)\sqrt{\chi}}{1-\sigma} + \frac{f}{b}\sqrt{\chi}(1+\sigma)2n(n-1)(n+1) + 2n(n-1)(n+1)(3-\sigma) \right]\\
&\ + J_{n+1}(x_-)J_{n+1}(x_+) \cdot 2(n^2-1)(1+\sigma)\sqrt{2}f\cdot |f|\frac{\sqrt{f^2-b^2}}{b}\cdot \frac{\sqrt{\chi}}{\sqrt{1-\sigma}} = 0,\\
\end{split}
\label{eq:master}
\end{equation} 
where we introduced the shorthand
\begin{eqnarray}
\chi \equiv 1+8\frac{b^2}{f^2}\cdot \frac{1-\sigma}{(1+	\sigma)^2}.
\end{eqnarray}
In the zero field case this equation reduces to the form derived by Sezawa \cite{sezawa1927dispersion}.


\section{Proof that $x_\pm \sim n$ for large $n$ at constant $s$}
\label{app:prooflinear}
Here we prove by contradiction that for $n\rightarrow \infty$ and $s$ fixed, both $x_+$ and $x_-$ scale proportionately to $n$. 
We begin with the proof for $x_+$. Assume first that $x_+ = O(n^\gamma)$ with $\gamma>1$. In the large $n$ limit eqs. \eqref{eq:poly1} and \eqref{eq:poly2} transform into 
\begin{eqnarray}
P_n(x_+,\alpha) &\approx& -J_n(x_+)x_+^2 \\
Q_n(x_+,\alpha) &\approx& J_n(x_+) \alpha^+ x_+^2.
\end{eqnarray}
Inserting this into eq. \eqref{eq:det} we find
\begin{eqnarray}
J_n(x_+) \approx 0.
\end{eqnarray}
Thus for large $n$ the $x_+$ are close to the roots of the Bessel functions. However, it is known \cite{Watson1995} that the $s$th root of the Bessel function $J_n(x)$ scales proportionately to $n$, thus $\gamma = 1$, yielding a contradiction. 

A similar type of argument shows that $\gamma < 1$ also leads to a contradiction. Proving that $x_+ = O(n)$. 

The same line of reasoning can also be applied for $x_-$ to prove $x_- = O(n)$. Thus we have shown that $x_\pm \sim n$ for large $n$ while holding $s$ fixed.

\section{Emergence of Rayleigh-waves in the $R\rightarrow \infty$ limit}
\label{app3}
Here we summarize the explanation by Viktorov \cite{viktorov1967rayleigh} how the $s=0$ mode becomes the Rayleigh edge-mode when $R\rightarrow \infty$. The argument relies on an asymptotic expansion of the Bessel functions $J_n(x)$ for large $n$ discovered by Debye \cite{debye1909naherungsformeln}, see also \cite{Watson1995}. The lowest term of this expansion states that
\begin{equation}
\label{eq:DebyeFormula}
J_n(x) \approx \frac{1}{\sqrt{2 \pi n \tanh \alpha}} \cdot e^{n(\tanh \alpha - \alpha)} 
\end{equation}
when $n$ is large, here $\cosh \alpha = n/x$. This asymptotic expression is valid as long as $x < n$. 
We can use this result to expand $J_n(q r)$ near the boundary. We first write $r=R-h$, with $h$ a positive number much smaller than $R$. We use the Debye formula \eqref{eq:DebyeFormula} and expand for small $h$. This yields the approximation
\begin{equation}
	J_n(q r) \sim e^{-\sqrt{n^2/R^2 - q^2 }(R-r)}
	\label{eq:approx2}
\end{equation}
valid if  $q R < n$ and $r<R$. Thus whenever the latter condition holds we have a mode exponentially localized near the boundary. This is precisely the Rayleigh mode. 

It can be checked numerically that for large enough $n$, the $s=0$ modes satisfy $q_+ r = x_+(s=0) < n$, thus $J_n(q_+ r)$ decays exponentially. The same is true for $J_n(q_- r)$, since $q_-$ is always imaginary. Therefore, at large enough $n$ the $s=0$ modes correspond to the Rayleigh-waves.

The modes of order $s\geq1$ do not satisfy the condition  $x_+(s) < n$, thus the argument cannot be applied to deduce exponential decay.

\section{Expression for the displacement fields}
\label{app2}

Using the definition of the displacement in terms of the potentials \eqref{eq:decomposition} and the general solution form \eqref{eq:solnOrder1}-\eqref{eq:solnOrder2} we find for the displacement fields

\begin{eqnarray}
\label{eq:displace}
u_r &=& \frac{1}{r}\bigg[ -A^+_n q_+ r J_{n+1}(q_+ r) +A^+_n n J_n(q_+ r)- A^-_n q_-r J_{n+1}(q_- r) + A^-_n n J_n(q_- r)\\ 
    && \phantom{nn}-n(\alpha^+ A^+_n J_n(q_+ r) + \alpha^- A^-_n J_n(q_- r) ) \bigg],   \nonumber\\
\label{eq:displacetheta}
u_\theta &=& \frac{i}{r} \bigg[n(A^+_n J_n(q_+ r/R) + A^-_n J_n(q_- r)) +\alpha^+ A^+_n q_+r J_{n+1}(q_+ r) - \alpha^+ A^+_n n J_n(q_+ r) \\
&& \phantom{nn} +\alpha^- q_-r A^-_n J_{n+1}(q_- r) -  \alpha^-A^-_n n J_n(q_- r)\bigg], \nonumber
\end{eqnarray}
where the ratio of amplitudes is determined by eq. \eqref{eq:det} and is given by
\begin{equation}
\begin{split}
&\ A^+_n[J_n(x_+) (-x^2_+ + (1-\sigma)(n-1)n(1-\alpha^+)) 
+ x_+J_{n+1}(x_+)(1-\sigma)(1+\alpha^+n)]  \\
&\  +A^-_n [J_n(x_-) (-x^2_- + (1-\sigma)(n-1)n(1-\alpha^-)) 
+ x_-J_{n+1}(x_-)(1-\sigma)(1+\alpha^-n)] =0. \\ 
\end{split}
\end{equation}		

Looking at equation \ref{eq:displace}, we see that $u_r$ is purely real, while $u_\theta$ is purely imaginary. 
Therefore, the trajectories of individual points of the medium describe ellipses, with radial and tangential directions corresponding to the major and minor axes.

\subsection{Relative contributions of the $+$ and $-$ branch in the cyclotron surface mode}
\label{app:relativecontrib}
In sec. \ref{sec:NewSurfModes} we explained how the condition for the localization of a mode near the disk boundary, is the smallness of the contribution of the $+$ part in the equations for the displacements \eqref{eq:displace} and \eqref{eq:displacetheta}. 
Using the explicit expressions for the amplitudes of the displacements we analyze here how much the $+$ and $-$ branches contribute to the full displacement.
We find numerically that the $-$ branch of the new-surface modes have no extrema within the disk. Thus these modes become maximal at the disk boundary.

We will denote the two contributions from the $+$ and $-$ branch in eqs. \eqref{eq:displace} and \eqref{eq:displacetheta} by $u_i(+)$ and $u_i(-)$, respectively. We find
		
\begin{eqnarray}
	 \frac{u_r(+)}{u_r(-)} &=& \frac{P(x_-,\alpha^-)}{P(x_+,\alpha^+)} \cdot \frac{n(1-\alpha^+) J_n(x_+) -x_+ J_{n+1}(x_+)  }{ n(1-\alpha^-) J_n(x_-) -x_- J_{n+1}(x_-) },  \\		 
	\frac{u_\theta(+)}{u_\theta(-)} &=& \frac{P(x_-,\alpha^-)}{P(x_+,\alpha^+)} \cdot \frac{n(1-\alpha ^+ )J_n(x_+) + x_+ \alpha^+ J_{n+1}(x_+)}{n(1-\alpha^- )J_n(x_-) + x_- \alpha^- J_{n+1}(x_-)},
\end{eqnarray}	
where we used eq. \eqref{eq:det} to express the ratios of $A^+$ and $A^-$ in terms of $P(x,\alpha)$. 
In the limit of $x_+ / n \gg 1$, using eq. \eqref{eq:poly1} we derive the asymptotics 
		
\begin{eqnarray}
	\label{eq:asymptoticsContribr}
	\frac{u_r(+)}{u_r(-)} &\propto & \frac{n}{x_+}, \\
	\frac{u_\theta (+)}{u_\theta (-)} & \propto & \frac{n}{x_+} 
	\label{eq:asymptoticsContribTheta}
\end{eqnarray}
by retaining the largest terms in the numerator and denominator. These asymptotics are valid when $J_n(x_+) \ne 0$. 

Thus we arrive at the explicit expressions for the displacement fields of the cyclotron surface modes
	
\begin{eqnarray}
	u_r &=& \frac{1}{r}A^-_n ( n(1-\alpha^-) J_n(q_-r) - q_- r J_{n+1}(q_-r)), 
	\label{eq:newmodes_r}\\
	u_\theta &=& \frac{i}{r} A^-_n (n(1-\alpha^-)J_n(q_-r) + q_- r \alpha^- J_{n+1}(q_-r)).
	\label{eq:newmodes_t}
\end{eqnarray}

\end{widetext}

\bibliography{biblio}

\end{document}